# Relaxation of a Quantum Emitter Resonantly Coupled to a Metal Nanoparticle

Khachatur V. Nerkararyan and Sergey I. Bozhevolnyi*

*Department of Technology and Innovation, University of Southern Denmark, Niels Bohrs Allé 1, DK-5230 Odense M, Denmark*

**Abstract**

Presence of a metal nanoparticle near a quantum dipole emitter, when a localized surface plasmon mode is excited via the resonant coupling with an excited quantum dipole, changes dramatically the relaxation dynamics: it is no longer described by an exponential decay but exhibits step-like behavior. The main physical consequence of this relaxation process is that the emission, being largely determined by the metal nanoparticle, comes out with a substantial delay. A large number of system parameters in our analytical description opens new possibilities for controlling quantum emitter dynamics.

*seib@iti.sdu.dk

Controlled modification of spontaneous emission is one of the central issues of quantum electrodynamics. Coupling between quantum dipole emitters (QDE) , such as molecules or quantum dots, and metal nanoparticles (MNP) at optical frequencies allows control over the flow of electromagnetic energy and lies at the core of an explosively growing field of quantum plasmonics [1]. Recent advances in nano-optics, especially experiments with single molecules interacting with well-defined metal nanostructures [2-4], often referred to as nano-antennas, serve as a strong impetus for further developments in this direction [5,6]. The most often discussed effect of QDE-MNP interaction is concerned with the modification (enhancement or quenching) of fluorescence yield determined by the balance between radiative and nonradiative decay rates, both enhanced near MNPs [3,4,7-9]. It is also expected that the QDE-MNP interaction can enter the regime of strong coupling, where excitation energy is coherently transferred between QDE and MNP in the form of Rabi oscillations [10].

Modification of spontaneous emission in various QDE-MNP configurations have been considered from different perspectives [3,4,7-9], but *always* assuming that the relaxation dynamics is *purely exponential* as obtained in the Weisskopf-Wigner treatment of an individual two-level atom [11]. In this Letter, we argue that, under *pulsed* excitation of the *resonantly* coupled QDE-MNP system, the relaxation dynamics changes drastically: it is no longer described by an exponential decay but exhibits step-like behavior. It should be noted that the most important feature of any QDE-MNP system is an extremely large difference between the relaxation times of the excited QDE and localized surface plasmon (LSP) mode of the MNP. Under these conditions and in the absence of the external illumination, it becomes crucial to properly take into account self-action of the excited QDE, in which its dipole field generates an LSP mode that acts back on the QDE thus providing a feedback in the QDE-MNP system. It turns out that, in the case of resonant QDE-MNP coupling, i.e., when the frequency of LSP resonance *coincides* with the frequency of the QDE radiative transition, the relaxation dynamics can be described *analytically* and features, in general, a step-like decay in time, deviating thereby *significantly* from the widely adopted exponential behavior [3,4,7-9]. It is further demonstrated that radiation emission, which is produced primarily by the MNP, comes out with a considerable time delay, opening thereby new possibilities for controlling the QDE-MNP emission process. We also discuss other possible realizations of the considered configuration.

The QDE-MNP system under consideration is schematically presented in Fig. 1, and consists of a generic three-level QDE [9, 10] and a spherical MNP. It is assumed that an external pump laser brings the QDE from the ground state 0 into the excited state 2, where it decays nonradiatively into the optically active state 1, and that the spherical MNP exhibits a dipolar LSP resonance at the frequency $\omega_{10}$ of the radiative (dipole-allowed) transition $1 \to 0$ [Fig. 1(b)]. This allows us to separate the excitation dynamics, which is not influenced by the presence of the MNP, from the relaxation dynamics of the state 1, whose modification due to the QDE-MNP coupling is the main subject of this work. Physically, a very similar situation can be realized with a two-level QDE under two-photon (pulsed) excitation. Note that the shape of a MNP is not important in this context and can be chosen specifically in order to produce a dipolar resonance at a given frequency [10,12], e.g., to coincide with the QDE radiative transition frequency.

We start our consideration when the QDE is brought (by a short pump pulse) into the optically active state 1, while the LSP is not excited (typical molecule relaxation times are anyway much long than the LSP lifetimes). We further assume that the QDE-MNP coupling is substantial, so that the QDE relaxation is determined by this coupling and not by interaction with vacuum fields as in the Weisskopf-Wigner treatment [11]. We anticipate thereby that the relaxation rate in our configuration will be significantly larger than the QDE relaxation rate $\gamma_0$ in free space. During the QDE relaxation, its wave function can be represented in the following form:

$$\Psi(t) = a_1(t)\phi_1 \exp\left(-\frac{i}{\hbar}E_1 t\right) + a_0(t)\phi_0 \exp\left(-\frac{i}{\hbar}E_0 t\right) \quad , \tag{1}$$

where $\phi_1$ and $\phi_0$ are the wave functions of the QDE states 1 and 0, characterized by the energies $E_1$ and $E_0$, respectively, while $a_1(t)$ and $a_0(t)$ are the corresponding probability amplitudes describing the transition $1 \to 0$. The QDE dipole moment is thereby given by:

$$\vec{D}(t) = a_1 a_0^* \vec{d_{10}} \exp\left(-i\omega_{10}t\right) + a_0 a_1^* \vec{d_{10}^*} \exp\left(i\omega_{10}t\right) \ , \tag{2}$$

with asterisk denoting the complex conjugate, $\vec{d_{10}} = \int \phi_1 e\vec{r}\phi_0^* dV$ and $\hbar\omega_{10} = E_1 - E_0$ being the dipole moment and energy of the transition $1 \to 0$.

Let us assume that the QDE dipole moment is collinear with the QDE-MNP axis and that the MNP center-to-QDE distance $R$ is considerably larger than the MNP radius $r$ [Fig. 1(a)], with all dimensions being much smaller than the wavelength $\lambda$ of light, i.e., that $\lambda \gg R \gg r$. In this electrostatic approximation, the MNP can be considered as being subjected to the homogenous electric field created by the oscillating QDE dipole. The LSP induced in the MNP by this field creates in its turn the electric field at the QDE site that can be written in the following form:

$$\vec{E_{sp}} = \frac{a_1 a_0^* \vec{d_1}}{\pi\varepsilon_0\varepsilon_2} \frac{\varepsilon_1 - \varepsilon_2}{\varepsilon_1 + 2\varepsilon_2} \frac{r^3}{R^6} \exp\left(-i\omega_{10}t\right) + \text{c.c.} \quad , \tag{3}$$

where c.c. stands for complex conjugate, $\varepsilon_0$ is the vacuum permittivity, $\varepsilon_1 = \varepsilon_{1r} + i\varepsilon_{1i}$ and $\varepsilon_2$ are the relative permittivities of the MNP and dielectric environment, respectively. In obtaining the above relation, we assumed that temporal variations of $a_1(t)$ and $a_0(t)$ are insignificant during the LSP lifetime, i.e., that the QDE-MNP dynamics is very slow in comparison with the LSP damping, an assumption that is consistent with the weak-coupling regime. Note that, if the QDE dipole moment is perpendicular to the QDE-MNP axis, the electric field given by Eq. (3) should be decreased by four times. Using the time-dependent Schrödinger equation for a two-level system in the driving field given by Eq. (3) and carrying out standard manipulations, one obtains the following system of two coupled equations for the probability amplitudes:

$$\frac{da_0}{dt} = \frac{i\left(\varepsilon_1^* - \varepsilon_2\right)\left|\overrightarrow{d_{10}}\right|^2 r^3}{\pi\hbar\varepsilon_0\varepsilon_2\left(\varepsilon_1^* + 2\varepsilon_2\right)R^6}a_1^*a_1a_0 \quad , \tag{4a}$$

$$\frac{da_1}{dt} = \frac{i\left(\varepsilon_1 - \varepsilon_2\right)\left|\overrightarrow{d_{10}}\right|^2 r^3}{\pi\hbar\varepsilon_0\varepsilon_2\left(\varepsilon_1 + 2\varepsilon_2\right)R^6}a_0^*a_0a_1 \quad . \tag{4b}$$

The obtained equations can be further simplified and made amenable to analytical treatment by considering the resonance configuration and relatively low LSP damping, i.e., with the following conditions being satisfied: $\left|\varepsilon_{1r} + 2\varepsilon_2\right| << \varepsilon_{1i}$ and $3\varepsilon_2 >> \varepsilon_{1i}$. In this case, the coupled equations become reduced to:

$$\frac{da_0}{dt} = \mu a_1^*a_1a_0 \quad \text{and} \quad \frac{da_1}{dt} = -\mu a_0^*a_0a_1 \quad , \tag{5}$$

with

$$\mu = \frac{3\left|\overrightarrow{d_{10}}\right|^2}{\pi\hbar\varepsilon_0\varepsilon_{1i}}\left(\frac{r}{R}\right)^3\frac{1}{R^3} \quad . \tag{6}$$

Note that Eq. (5) implies that $\left|a_0\right|^2 + \left|a_1\right|^2 = const$. Imposing the initial conditions, $a_0(\tau) = a_{00}$ and $a_1(\tau) = a_{10} = \sqrt{1 - a_{00}^2}$, allows us to write the solution:

$$a_0(t - \tau) = \frac{a_{00}}{\sqrt{a_{00}^2 + a_{10}^2\exp\left[-2\mu(t - \tau)\right]}} \quad , \quad a_1(t - \tau) = \frac{a_{10}}{\sqrt{a_{10}^2 + a_{00}^2\exp\left[-2\mu(t - \tau)\right]}} \quad . \tag{7}$$

The QDE relaxation dynamics is strongly influenced by the excitation of the LSP mode that opens a very efficient relaxation channel. This channel is primarily nonradiative, but the MNP dipole moment does contribute to the emission process, dominating in fact over the QDE radiation. The total dipole moment of the QDE-MNP system can be represented in the following form:

$$\vec{P} = \left(1 - \frac{6i\varepsilon_2}{\varepsilon_{1i}} \frac{r^3}{R^3}\right) \frac{\overline{d_{10}} \exp\left(i\omega_{10}t\right)}{2\cosh\left[\mu\left(t-\tau\right) - \ln\left(a_{10}/a_{00}\right)\right]} + \text{c.c.} \quad .$$  (8)

The above expressions [Eq. (6) - (8)] constitute the main theoretical result of our work, providing simple analytical formulae for the QDE-MNP relaxation and emission dynamics and demonstrating that the dynamics is in general quite complicated. The value of parameter $\mu$ determines the characteristic relaxation rate of the excited QDE state. The QDE relaxation described by Eq. (7) begins at some time moment $\tau$ when the QDE state 1, which is created by nonradiative decay of the excited state 2, is partially relaxed into the ground state 0, so that $a_{00} > 0$. This starting process can occur due other inducements always found in an open system, for example, due to the free-space spontaneous emission, i.e., without the QDE-MNP dipole-dipole interaction because the QDE dipole moment [Eq. (2)] is still negligibly small.

One of the most important assumptions made is related to the strength of the QE-MNP coupling which should ensure considerably larger relaxation rates $\mu$ than that for the QE in free space ($\gamma_0$). Their ratio can be evaluated now with the help of Eq. (6) and the Weisskopf-Wigner result [11] as follows

$$\beta \cong \frac{\mu}{\gamma_0} = \frac{9}{\varepsilon_{1i}\sqrt{\varepsilon_2}} \left(\frac{\lambda_0}{2\pi R}\right)^3 \left(\frac{r}{R}\right)^3 \quad ,$$  (9)

with $\lambda_0$ being the vacuum wavelength corresponding to the QDE transition frequency $\omega_{10}$. For a typical dielectric environment with $\varepsilon_2 = 2.25$ (e.g., glass or polymer), the resonance condition (i.e., $\varepsilon_{1r} = -4.5$) is met, for gold, at the wavelength of ~ 530 nm with $\varepsilon_{1i}^g \cong 2.35$ and, for silver, at ~ 400 nm with $\varepsilon_{1i}^s \cong 0.22$ [13]. Considering an MNP with the radius of 5 nm and the QDE distance to the MNP center being 15 nm (in order to be within the electrostatic dipole description), one obtains the ratio $\beta \approx 17$ for gold and $\approx 77$ for silver, justifying thereby the above assumption, $\mu >> \gamma_0$. It is interesting that the effect is already pronounced at relatively large (~ 10 nm) distances between QDEs and the MNP surface, which are in the range of distances explored in the recent experiments with 10-nm-size gold nanoparticles [8]. It is also transparent that even larger ratios can be achieved by exploiting the LSP shape dependence [10,12] and red-shifting the MNP resonance towards smaller metal absorption [13].

The QDE relaxation described by Eq. (7) starts off when nonzero population of the ground state is reached due to other (relatively slow) relaxation processes with an exponential decay, so that $a_1^r(t) = \exp(-\gamma_0 t)$. Applying the continuity condition at the transition between these two processes to both functions, $a_1^r(t)$ and $a_1(t)$, and their derivatives, one can determine the characteristic time $\tau = 1/2\mu$. Note that this time does not depend on the QDE relaxation rate $\gamma_0$ in free space. Therefore, the present consideration allows one to analyze the whole relaxation process. The QDE relaxation dynamics depends strongly on the efficiency of the QDE-MNP interaction, which is characterized by the relaxation rate ratio $\beta$ [Eq. (9)]. Sharp step-like behavior observed for very large values of $\beta$ changes to more gradual population decay for smaller $\beta$ [Fig. 2(a)]. Spontaneous emission from the QDE-MNP system is determined by the square magnitude of the system dipole moment [Eq. (8)], whose maximum is attained after a certain delay time $t_d = [1 + \ln(2\beta)]/(2\beta\gamma_0)$. For strong QDE-MNP interactions (very large $\beta$), the emission peak is narrow and occurs close to the initial moment of time [Fig. 2(b)]. It should also be noted that the delay time $t_d$ is of the same order of magnitude as the width (at half maximum) of the emission peak. This interesting feature might be found useful when conducting and analyzing the corresponding experiments.

We would like to mention that the occurrence of transient effects at early times was noted in the theoretical consideration of molecular dynamics modified by the presence of an MNP, stressing the following (at later times) exponential decay in the case of weak coupling but without further analysis of transient behavior [10]. Numerical analysis based on the SPP quantum-mechanical description indicated however the occurrence of the emission peak with a certain delay in time in a fashion similar to our results (cf. insets in Fig. 2 here and Fig. 2 in [10]). Finally, we believe that such a delay has actually been present (but not noted) in the recent experiments with 10-nm-size gold nanoparticles connected by DNA to individual fluorophores (see Fig. 1(d) in [8]).

The most interesting physical finding of our work, viz., the spontaneous emission delay under pulsed excitation, is in fact quite general. Indeed, it is only required that a QDE is placed near a resonator that, at the frequency of QDE radiative transition, features a well-defined dipolar resonance with the damping rate, which is substantially larger than the QDE relaxation rate in free space. The former ensures the $\pi/2$ phase delay in its electromagnetic feedback, which is essential for arriving at Eq. (5), while the latter is needed to realize the desirable weak-coupling regime. The appropriate interaction can be realized at practically any wavelength with non-spherical MNPs [12] or by using low absorbing dielectric (semiconductor) nanoparticles having large permittivity values and supporting strong Mie resonances that can be chosen propitiously by adjusting the particle shapes and sizes [14]. Another possibility would be to place a QDE near a metal surface, a configuration that is resonant if $\varepsilon_{1r}(\omega_{10}) = -\varepsilon_2$, with the strong-coupling regime requiring sub-nanometer QDE-surface distances [15]. Our approach can also be applied in this case, provided that $|\varepsilon_{1r} + \varepsilon_2| << \varepsilon_{1i}$ and $2\varepsilon_2 >> \varepsilon_{1i}$, resulting in similar emission dynamics with the relaxation parameter given by

$$\mu_1 = \frac{\left|\overrightarrow{d_{10}}\right|^2}{8\pi\hbar\varepsilon_0\varepsilon_{1i}R^3} \quad , \tag{10}$$

with $R$ being in this case the QDE-surface distance and considering that the QDE dipole moment is perpendicular to the metal surface. Note, that the $R^{-6}$ scaling in Eq. (6) is transformed, for this configuration, into the $R^{-3}$ scaling [Eq. (10)], which is also expected to be the case for small QE-MNP separations with the dipolar MNP response to the homogeneous field becoming strongly multipolar and approaching that of a flat metal surface.

In summary, we have considered the relaxation dynamics of a generic QDE excited with short pump pulses and located near a MNP that exhibits a dipolar LSP resonance at the frequency of the QDE radiative transition. Our theoretical analysis resulted in the following conclusions: (*i*) the relaxation dynamics in the resonantly coupled QDE-MNP system exhibits step-like behavior deviating thereby significantly from the generally accepted exponential decay [3,4,7-9], (*ii*) the QDE-MNP radiation emission reaches its maximum with a significant delay in time, and (*iii*) a large number of system parameters in our analytical description opens new possibilities for controlling the QDE relaxation and emission dynamics. Given the variety of resonant plasmonic [12] and semiconductor [14] nanoparticles, the experimental observation of the predicted effect seems feasible, while the possibility of tuning the delay time by changing the QDE-MNP separation can be exploited in many applications, e.g., for realizing a nanoscopic ruler [16], as well as in fundamental studies within quantum plasmonics [1].

We acknowledge financial support for this work from the VELUX Foundation and from the Danish Council for Independent Research (the FTP project ANAP, Contract No.09-072949).

**Figure captions**

FIG. 1 (color online). Schematic of a system with a quantum dipole emitter (QDE) placed near a metal nanoparticle (MNP), indicating (a) system parameters and (b) QDE energetic levels along with the localized (dipolar) surface plasmon resonance of the MNP.

FIG. 2 (color online). Relaxation dynamics of the QDE-MNP configuration for different ratios of relaxation rates $\beta$ [Eq. (9)], showing (a) population decay of the excited state 1 [Eq. (7)] and (b) the squared magnitude of the total dipole moment [Eq. (8)] and normalized to its maximum value. Insets display the same dependencies for $\beta = 20$ in the logarithmic scale.

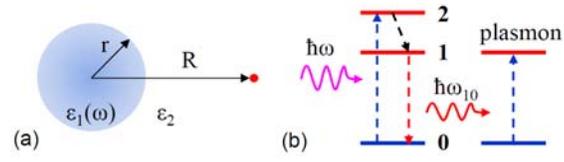

Fig. 1

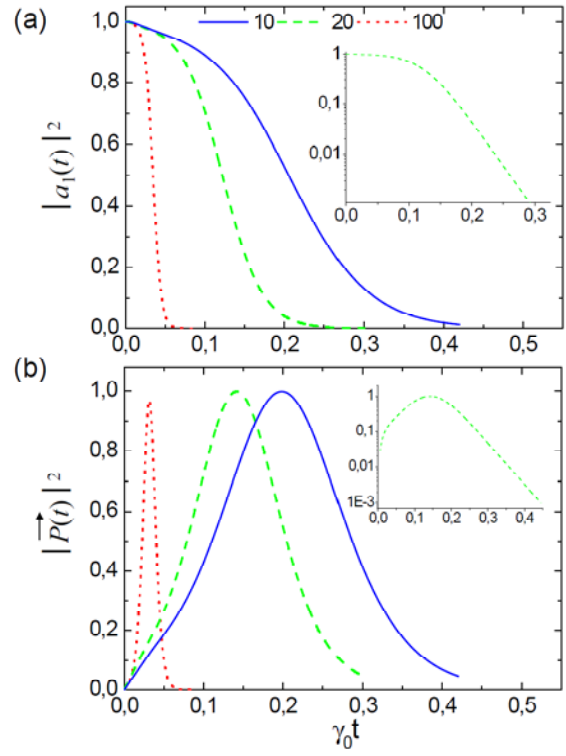

Fig. 2